\documentstyle[aps,prl,preprint]{revtex}

\begin{document}
\title{Comment on ``Manipulating the frequency-entangled states by an
acoustic-optical modulator''}
\author{K.J.Resch, S.H.Myrskog, J.S.Lundeen and A.M.Steinberg}
\address{Department of Physics, University of Toronto\\
Toronto, ON, M5S\ 1A7\\
CANADA}
\maketitle

\begin{abstract}
A recent theoretical paper [1] proposes a scheme for entanglement swapping
utilizing acousto-optic modulators without requiring a Bell-state
measurement. \ In this comment, we show that the proposal is flawed and no
entanglement swapping can occur without measurement.
\end{abstract}

\newpage \tightenlines

\section{Introduction}

\bigskip

\bigskip

Entanglement swapping, a term coined by \cite{thentswap}, is the process of
creating entanglement between two particles that have never before
interacted. \ When one generates entangled states, a pair of particles are
created in tandem. \ The most common processes used to generate these
entangled pairs are atomic cascades \cite{cascade} and spontaneous
parametric downconversion \cite{mandel, kwiat}. \ In an entanglement
swapping scheme, one begins with a {\em pair} of 2-particle entangled pairs.
\ In successful schemes to date \cite{expentswap}, one performs a Bell
measurement on two of the particles -- one from each entangled pair. \ A
successful Bell-state measurement collapses the remaining particles into a
new entangled state - even though the particles have not directly
interacted. \ The process of entanglement swapping was central to the
experimental realization of quantum teleportation \cite{teleportation}. \ 

This comment is on a recent proposal \cite{AOMpaper} for performing
entanglement swapping with acousto-optic modulators (AOMs) without requiring
a Bell measurement. \ The authors make a faulty assumption about the
transformation an AOM\ performs on its input photon modes, which leads to
incorrect conclusions. \ In this comment, we describe generally how one
should treat the interaction of an AOM\ with its two input light fields
quantum mechanically. \ Then we apply this type of interaction to the
proposed scheme, and show that no entanglement swapping can take place.

\bigskip

\section{Theory}

\bigskip

\subsection{General theory}

An acousto-optic modulator can be used to couple two modes of an
electromagnetic field by means of a phonon field. \ A simple diagram from
[1] shows this schematically (figure 1). \ The interaction between two input
fields and an acousto-optic modulator will be described by an effective
Hamiltonian, ${\cal H}_{{\rm eff}}$, of the form: 
\begin{equation}
{\cal H}_{{\rm eff}}=gb(\delta )a_{t}(\omega )a_{d}^{\dagger }(\omega
+\delta )+g^{\ast }b^{\dagger }(\delta )a_{t}^{\dagger }(\omega
)a_{d}(\omega +\delta ),
\end{equation}
where $b$ and $b^{\dagger }$ are the annihilation and creation operators for
the phonon field, $a$ and $a^{\dagger }$ are the annihilation and creation
operators for the photon field, and g is the coupling constant. \ The
subscripts $t$ and $d$ refer to the mode labels shown in figure 1, and $%
\omega $ and $\omega +\delta $ are the photon frequencies. \ The first term
in this Hamiltonian describes the destruction of a phonon of frequency $%
\delta $, and a photon of frequency $\omega $ in mode d, and the creation of
a photon with frequency $\omega +\delta $ in mode t. \ The second term in
the Hamiltonian describes the creation of a phonon of frequency $\delta $,
and a photon of frequency $\omega $ in mode d, and the destruction of a
photon in mode t. \ This Hamiltonian is manifestly hermitian, and the
propagator that follows from it must be unitary. \ If one assumes that the
phonon field in the AOM\ is a classical field, which is a reasonable
approximation for a coherent state of phonons with a high average phonon
number, then we can replace the phonon operators with c-numbers $\beta $ and 
$\beta ^{\ast }$. \ The Hamiltonian then becomes: 
\begin{equation}
{\cal H}_{{\rm eff}}=g\beta a_{t}(\omega )a_{d}^{\dagger }(\omega +\delta
)+g^{\ast }\beta ^{\ast }a_{t}^{\dagger }(\omega )a_{d}(\omega +\delta ).
\end{equation}

Over an inifinitesimal interaction time, $dt$, the AOM will perform the
following transformations: 
\begin{eqnarray}
\left| \omega \right\rangle _{1} &\rightarrow &\left| \omega \right\rangle
_{t}-\frac{i}{\hbar }g\beta \left| \omega +\delta \right\rangle _{d}dt 
\nonumber \\
\left| \omega +\delta \right\rangle _{1^{\prime }} &\rightarrow &\left|
\omega +\delta \right\rangle _{d}-\frac{i}{\hbar }g^{\ast }\beta ^{\ast
}\left| \omega \right\rangle _{t}dt.
\end{eqnarray}
Over longer times, the transformation becomes: 
\begin{eqnarray}
\left| \omega \right\rangle _{1} &\rightarrow &\cos \left( \frac{\left|
g\beta \right| t}{\hbar }\right) \left| \omega \right\rangle _{t}-i\frac{%
g\beta }{\left| g\beta \right| }\sin \left( \frac{\left| g\beta \right| t}{%
\hbar }\right) \left| \omega +\delta \right\rangle _{d}  \nonumber \\
\left| \omega +\delta \right\rangle _{1^{\prime }} &\rightarrow &\cos \left( 
\frac{\left| g\beta \right| t}{\hbar }\right) \left| \omega +\delta
\right\rangle _{d}-i\frac{g^{\ast }\beta ^{\ast }}{\left| g\beta \right| }%
\sin \left( \frac{\left| g\beta \right| t}{\hbar }\right) \left| \omega
\right\rangle _{t}.
\end{eqnarray}
We can choose the interaction time to create equal superpositions of the
outgoing modes and define the phase angle, $\phi =\arg (g\beta ).$ The
transformations then become: 
\begin{eqnarray}
\left| \omega \right\rangle _{1} &\rightarrow &\frac{1}{\sqrt{2}}\left[
\left| \omega \right\rangle _{t}-ie^{i\phi }\left| \omega +\delta
\right\rangle _{d}\right]  \nonumber \\
\left| \omega +\delta \right\rangle _{1^{\prime }} &\rightarrow &\frac{1}{%
\sqrt{2}}\left[ \left| \omega +\delta \right\rangle _{d}-ie^{-i\phi }\left|
\omega \right\rangle _{t}\right] .  \label{goodtransforms}
\end{eqnarray}

\bigskip

\subsection{\protect\bigskip An AOM cannot result in entanglement swapping
without a Bell measurement}

The authors of \cite{AOMpaper} claim that an AOM (figure 1) can be modelled
by taking 2 input modes, 1 and 1$^{^{\prime }}$, and transforming them to
two output modes as follows: 
\begin{eqnarray}
&&\left| \omega \right\rangle _{1}\stackrel{AOM}{\rightarrow }\frac{1}{\sqrt{%
2}}\left[ \left| \omega \right\rangle _{t}+\left| \omega +\delta
\right\rangle _{d}\right]  \nonumber \\
&&\left| \omega +\delta \right\rangle _{1^{\prime }}\stackrel{AOM}{%
\rightarrow }\frac{1}{\sqrt{2}}\left[ \left| \omega \right\rangle
_{t}+\left| \omega +\delta \right\rangle _{d}\right] .  \label{nonunitary}
\end{eqnarray}
\ The equations above actually differ from equations (1) and (2) from \cite
{AOMpaper} due to a presumed typographical error in the left side of the
second equation, but are consistent with the rest of their paper. \ However,
such a transform is not allowed by quantum mechanics as it is non-unitary. \
In other words, the two input states are orthogonal, and must remain so by
any unitary transformation. \ As one can see from the proposed
transformation, the final states are not orthogonal - in fact they are
identical. \ Such transformations destroy information and lead to paradoxes
such as superluminal signalling. \ In the present case, if the proposed
scheme were correct, a decision by Alice of whether or not to perform the
transformation could instantaneously affect a measurement by Bob of whether
or not his photon pair was entangled.

Instead of the transformation given in \cite{AOMpaper}, one should model the
AOM by the unitary transformation described previously. \ We use the
transforms from equation \ref{goodtransforms} and make the assumption that $%
\phi =0$, without loss of generality. \ To put the transform into the same
form as equation \ref{nonunitary}, the second transform is multiplied by a
phase of $\exp (i\pi /2)$: 
\begin{eqnarray}
\left| \omega \right\rangle _{1} &\rightarrow &\frac{1}{\sqrt{2}}\left[
\left| \omega \right\rangle _{t}-i\left| \omega +\delta \right\rangle
_{d}\right]  \nonumber \\
\left| \omega +\delta \right\rangle _{1^{\prime }} &\rightarrow &\frac{1}{%
\sqrt{2}}\left[ \left| \omega \right\rangle _{t}+i\left| \omega +\delta
\right\rangle _{d}\right] .
\end{eqnarray}
The negative sign in the first term ensures that the final states remain
orthogonal, preserving angle in the 2-dimensional Hilbert space. \ We now
follow through the calculations from \cite{AOMpaper} and describe the
separate 2-particle entangled states (see figure 2) as: 
\begin{eqnarray}
\left| \phi \right\rangle &=&\frac{1}{\sqrt{2}}\left[ \left| \omega
\right\rangle _{1}\left| \omega +\delta \right\rangle _{2}+\left| \omega
+\delta \right\rangle _{1^{\prime }}\left| \omega \right\rangle _{2^{\prime
}}\right]  \nonumber \\
\left| \psi \right\rangle &=&\frac{1}{\sqrt{2}}\left[ \left| \omega
\right\rangle _{3}\left| \omega +\delta \right\rangle _{4}+\left| \omega
+\delta \right\rangle _{3^{\prime }}\left| \omega \right\rangle _{4^{\prime
}}\right] .
\end{eqnarray}
The states $\left| \phi \right\rangle $ and $\left| \psi \right\rangle $
refer to the states of the particles created at the entangled-photon sources
1 and 2 respectively. \ The primed and unprimed subscripts refer to the
spatial modes of the photons (figure 2), and the labels $\omega $ and $%
\omega +\delta $ refer to their angular frequencies. \ The two photons
described in these states are not only entangled in their energy (frequency)
but also in their spatial paths. \ We can now apply the following unitary
transformations to the modes that interact with the AOMs in the scheme: \ 
\begin{eqnarray}
&&\left| \omega +\delta \right\rangle _{2}\stackrel{AOM1}{\rightarrow }\frac{%
1}{\sqrt{2}}\left[ \left| \omega \right\rangle _{T_{1^{\prime }}}+i\left|
\omega +\delta \right\rangle _{T_{1}}\right]  \nonumber \\
&&\left| \omega \right\rangle _{3}\stackrel{AOM1}{\rightarrow }\frac{1}{%
\sqrt{2}}\left[ \left| \omega \right\rangle _{T_{1^{\prime }}}-i\left|
\omega +\delta \right\rangle _{T_{1}}\right]  \nonumber \\
&&\left| \omega +\delta \right\rangle _{3^{\prime }}\stackrel{AOM2}{%
\rightarrow }\frac{1}{\sqrt{2}}\left[ \left| \omega \right\rangle
_{T_{2}}+i\left| \omega +\delta \right\rangle _{T_{2^{\prime }}}\right] 
\nonumber \\
&&\left| \omega \right\rangle _{2^{\prime }}\stackrel{AOM2}{\rightarrow }%
\frac{1}{\sqrt{2}}\left[ \left| \omega \right\rangle _{T_{2}}-i\left| \omega
+\delta \right\rangle _{T_{2^{\prime }}}\right] .
\end{eqnarray}
AOM1 and AOM2 simply refer to the transformation applied by the AOMs marked
1 and 2 in figure 2. \ 

Using these transformations, the inital state describing the four photons, $%
\left| \phi \right\rangle \otimes \left| \psi \right\rangle $, will become:
\ 
\begin{eqnarray}
\left| \phi \right\rangle \otimes \left| \psi \right\rangle  &=&\frac{1}{2}%
\left\{ 
\begin{array}{c}
\left[ \left| \omega \right\rangle _{1}\left( \left| \omega \right\rangle
_{T_{1^{\prime }}}+i\left| \omega +\delta \right\rangle _{T_{1}}\right)
+\left| \omega +\delta \right\rangle _{1^{\prime }}\left( \left| \omega
\right\rangle _{T_{2}}-i\left| \omega +\delta \right\rangle _{T_{2^{\prime
}}}\right) \right] \otimes  \\ 
\left[ \left| \omega +\delta \right\rangle _{4}\left( \left| \omega
\right\rangle _{T_{1^{\prime }}}-i\left| \omega +\delta \right\rangle
_{T_{1}}\right) +\left| \omega \right\rangle _{4^{\prime }}\left( \left|
\omega \right\rangle _{T_{2}}+i\left| \omega +\delta \right\rangle
_{T_{2^{\prime }}}\right) \right] 
\end{array}
\right\}  \\
&=&\frac{1}{2}\left\{ 
\begin{array}{c}
\left| \omega \right\rangle _{1}\left| \omega +\delta \right\rangle
_{4}\left( \left| \omega \right\rangle _{T_{1^{^{\prime }}}}+i\left| \omega
+\delta \right\rangle _{T_{1}}\right) \left( \left| \omega \right\rangle
_{T_{1^{^{\prime }}}}-i\left| \omega +\delta \right\rangle _{T_{1}}\right) +
\\ 
\left| \omega +\delta \right\rangle _{1^{\prime }}\left| \omega +\delta
\right\rangle _{4}\left( \left| \omega \right\rangle _{T_{2}}-i\left| \omega
+\delta \right\rangle _{T_{2^{\prime }}}\right) \left( \left| \omega
\right\rangle _{T_{1^{^{\prime }}}}-i\left| \omega +\delta \right\rangle
_{T_{1}}\right) + \\ 
\left| \omega \right\rangle _{1}\left| \omega \right\rangle _{4^{\prime
}}\left( \left| \omega \right\rangle _{T_{1^{^{\prime }}}}+i\left| \omega
+\delta \right\rangle _{T_{1}}\right) \left( \left| \omega \right\rangle
_{T_{2}}+i\left| \omega +\delta \right\rangle _{T_{2^{\prime }}}\right) + \\ 
\left| \omega +\delta \right\rangle _{1^{\prime }}\left| \omega
\right\rangle _{4^{\prime }}\left( \left| \omega \right\rangle
_{T_{2}}-i\left| \omega +\delta \right\rangle _{T_{2^{\prime }}}\right)
\left( \left| \omega \right\rangle _{T_{2}}+i\left| \omega +\delta
\right\rangle _{T_{2^{\prime }}}\right) 
\end{array}
\right\} .
\end{eqnarray}
The authors propose to discard the cases where both photons go through the
same AOM (the first and fourth terms in the above equation) and are left
with only the remaining 2 terms. \ These terms are: 
\begin{eqnarray}
&&\left| \omega +\delta \right\rangle _{1^{\prime }}\left| \omega +\delta
\right\rangle _{4}\left( \left| \omega \right\rangle _{T_{2}}-i\left| \omega
+\delta \right\rangle _{T_{2^{\prime }}}\right) \left( \left| \omega
\right\rangle _{T_{1^{^{\prime }}}}-i\left| \omega +\delta \right\rangle
_{T_{1}}\right) + \\
&&\left| \omega \right\rangle _{1}\left| \omega \right\rangle _{4^{\prime
}}\left( \left| \omega \right\rangle _{T_{1^{^{\prime }}}}+i\left| \omega
+\delta \right\rangle _{T_{1}}\right) \left( \left| \omega \right\rangle
_{T_{2}}+i\left| \omega +\delta \right\rangle _{T_{2^{\prime }}}\right) . 
\nonumber
\end{eqnarray}
It is apparent that when the proper transformation is used, the terms
describing the light after the AOM do {\em not} factor out and no
entanglement swapping has occurred between photons 1 and 4. \ In fact, the
states describing the light after passing through the AOM have an overlap of
zero, precluding even partial entanglement swapping.\ 

In \cite{AOMpaper}, a different AOM scheme is used to create a
Greenberger-Horne-Zeilinger (GHZ) 3-particle entangled state using a pair of
2-photon entangled states. \ Unfortunately, the same tranformation as shown
in equation (1) is used to model the AOM. \ When the appropriate
transformation is applied instead to their scheme, there is no 3-particle
entanglement.

\section{Conclusion}

\bigskip

The authors of \cite{AOMpaper} used a non-unitary transformation to describe
the action of an AOM\ on a pair of input photon modes, and this appeared to
lead to unconditional entanglement swapping. \ We have shown that when a
unitary transformation is used instead, as required by quantum mechanics, no
entanglement between the photons from different sources is achieved. \ Due
to the same erroneous transformation, the claim that an AOM\ could create a
GHZ\ state using a pair of 2-photon entangled states is also incorrect. \ In
general, no unitary transformation on one pair of photons can ever modify
the reduced density matrix of a different pair. \ This is why effects such
as entanglement swapping \cite{expentswap} and quantum teleportation \cite
{teleportation} always require a nonunitary interaction (measurement).

We acknowledge the financial support of NSERC\ and Photonics Research
Ontario. \ S.H. Myrskog and K.J. Resch also acknowledge the support of the
Walter C. Sumner foundation.

Figure 1. \ \ The two input modes, 1 and 1$^{\prime }$, enter an AOM and are
converted to two output modes, $t$ and $d$.

\bigskip

Figure2. \ The schematic for the proposed entanglement swapping scheme.

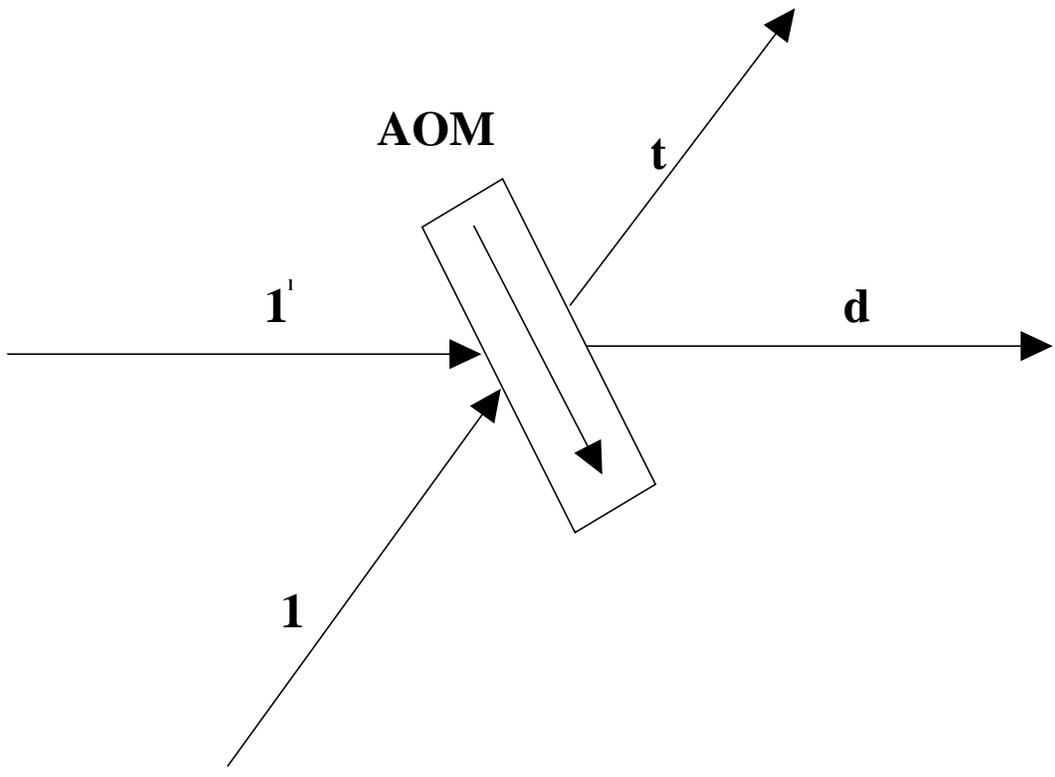
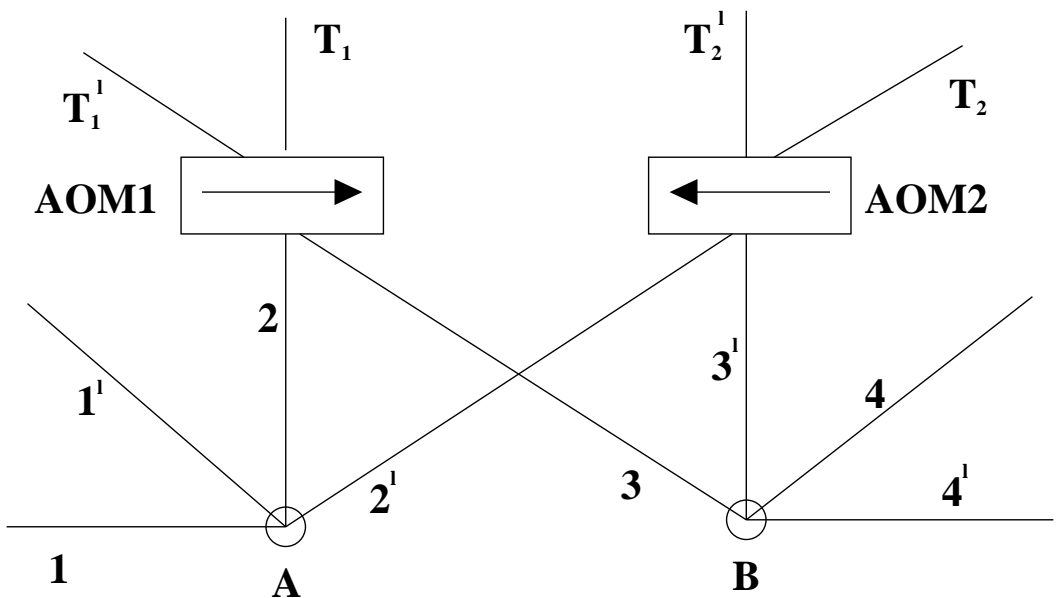
\end{document}